\documentclass[aps,showpacs,prl,twocolumn]{revtex4}
\usepackage{mathrsfs}
\usepackage{amssymb}
\usepackage{amsmath}
\usepackage{bm}
\usepackage{graphics}
\usepackage{natbib}

\begin{document}
\title{Majorana modes in a triple-terminal Josephson junction with embedded parallel-coupled double quantum dots}
\author{Guang-Yu Yi}
\author{Xiao-Qi Wang}
\author{Zhen Gao}
\author{Hai-Na Wu}
\author{Wei-Jiang Gong}\email{gwj@mail.neu.edu.cn}

\affiliation{College of Sciences, Northeastern University, Shenyang
110819, China}

\date{\today}

\begin{abstract}
We investigate the Josephson effect in one triple-terminal junction with embedded parallel-coupled double quantum dots.
It is found that the inter-superconductor supercurrent has opportunities to oscillate in $4\pi$ period, with the adjustment of the phase differences among the superconductors. What is notable is that such a result is robust and independent of fermion parities, intradot Coulomb strength, and the dot-superconductor coupling manner. By introducing the concept of spinful many-particle Majorana modes, we present the analytical definition of the Majorana operator via superposing electron and hole operators. It can be believed that this work provide a simple but feasible proposal for the realization of Majorana modes in a nonmagnetic system.
\end{abstract}
 \maketitle

\bigskip

Majorana fermions, exotic quasiparticles with non-Abelian statistics, have attracted enormous attention because of their fundamental physics and potential application in topological quantum computation. During the past years, various schemes have been proposed to realize unpaired Majorana fermions, such as in a vortex core in a $p$-wave superconductor [1-5] or superfluid [6-8]. Feasible proposals are to fabricate Majorana bound states (MBSs) at the ends of a one-dimensional $p$-wave superconductor, which has been constructed by adhering a Rashba nanowire with a strong magnetic field or a ferromagnetic chain to the $s$-wave superconductor [9-14]. However, how to verify the existence of MBSs is a key issue and is rather difficult.

\begin{figure}[htb]
\begin{center}\scalebox{0.29}{\includegraphics{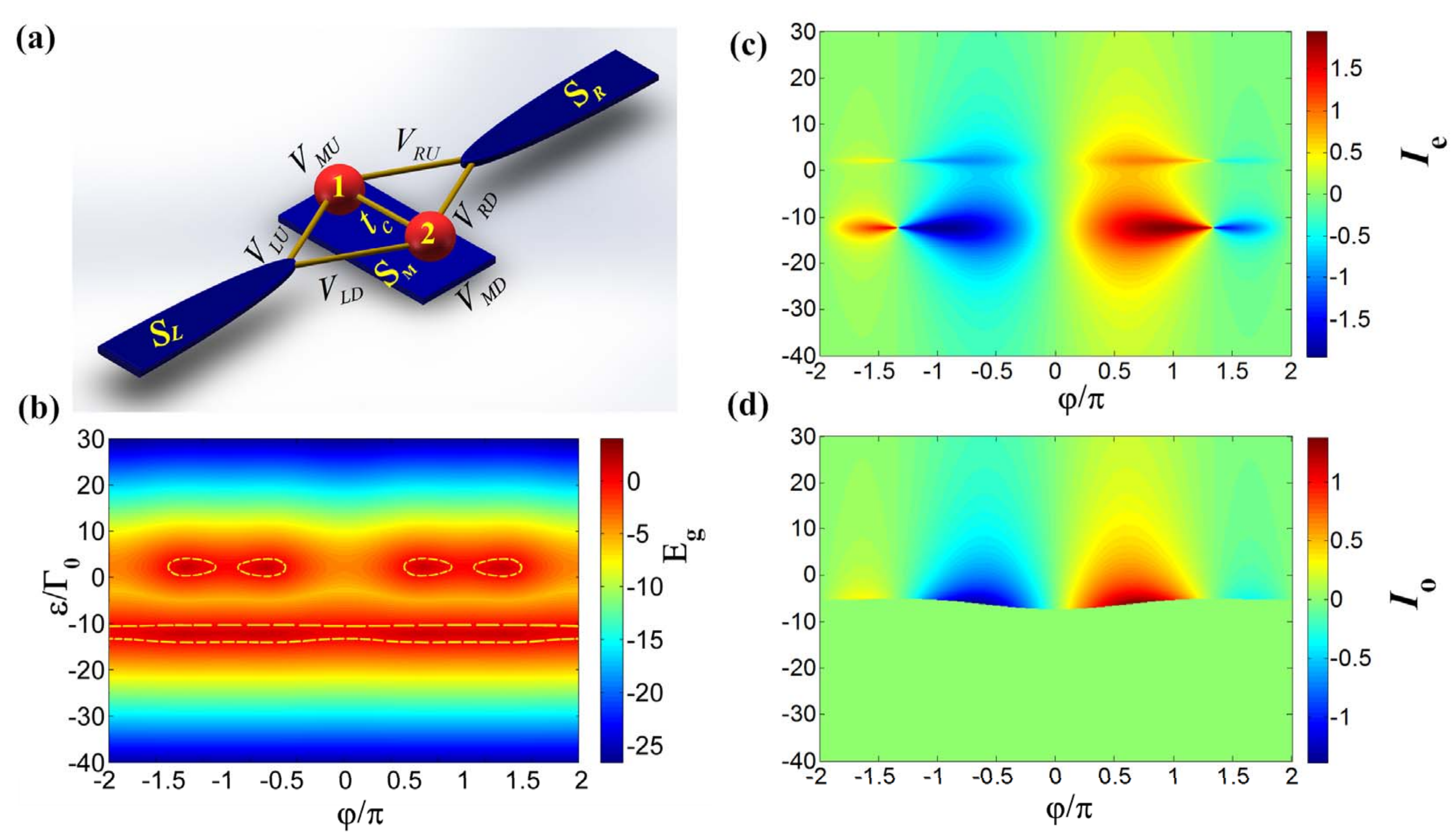}}
\caption{(a) Schematic of a Josephson junction with a DQD molecule. (b) Difference between the GS levels in even and odd FPs.
(c)-(d) Josephson currents in the even and odd FPs, respectively.
\label{Struct}}
\end{center}
\end{figure}
Superconductor-quantum-dot systems are also proposed to realize Majorana modes, by including
spin-orbit coupling or anisotropic magnetic fields [15-17]. Although quantum dots (QDs) have several
advantages, strong spin-orbit or anisotropic magnetic fields form a major hurdle and limit material flexibility.
For example, anisotropic magnetic fields only result in spinless localized Majorana modes when $E_Z\gg \Delta\sim t$ with $\Delta$ the superconducting gap and $t$ the intersite hopping [17]. In order to overcome such difficulty, the structure of superconductor/double-QD/superconductor (SC/DQD/SC) has recently been designed [18]. It has been observed that in such a structure, spinful many-particle Majorana modes can arise in the presence of appropriate Coulomb interactions and local magnetic fields applied on the QDs. However, for practical experiments, it is a formidable challenge to apply two different local magnetic fields on the nanoscale. Therefore, any new suggestion to realize the Majorana modes in the non-magnetic systems is desirable.
\par
Some recent researches suggest that a triple-terminal Josephson junction can be a promising candidate for achieving Majorana modes. Main reason is that in this junction, the Kramers
degeneracy of its different-parity energy levels can be broken in the presence of superconducting phase difference, and adjusting phase
differences can change the ground-state (GS) parity [19]. Motivated by these phenomena, in this Letter we would like to investigate the Josephson effect in one triple-terminal junction with embedded parallel-coupled DQDs. As a result, by adjusting the phase differences among the superconductors, the $4\pi$-periodic oscillation of the inter-SC supercurrent appears, which is independent of structural parameters and fermion parities. Then, we define the Majorana operator via superposing electron and hole operators. Hence, this work provides a simple but feasible proposal to realize Majorana modes in non-magnetic systems.

\begin{figure}[htb]
\begin{center}\scalebox{0.45}{\includegraphics{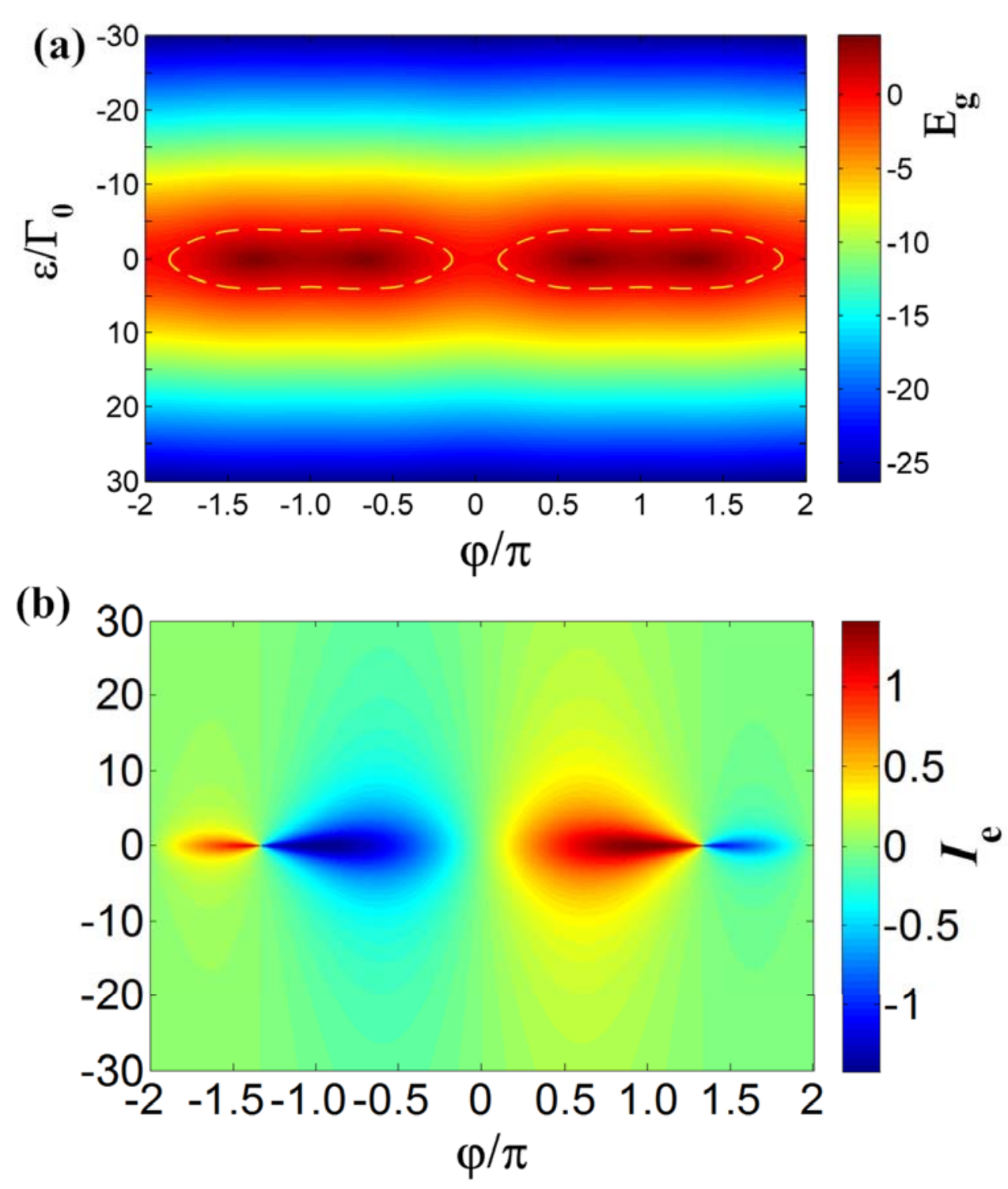}}
\caption{(a) Difference between the GS levels in even and odd FPs in the case of infinite Coulomb interaction.
(b) Corresponding Josephson current in the even-FP case.
\label{Struct}}
\end{center}
\end{figure}
The considered Josephson junction is illustrated in Fig.1(a), where parallel-coupled DQDs couple to three SCs, respectively. The Hamiltonian for the electron motion in this system can be written as
$H=\sum_{\alpha=L,M,R}H_\alpha+H_D+H_T$, and
\begin{small}
\begin{eqnarray}
H_\alpha&=&\sum_{k\sigma}\varepsilon_{\alpha k} a^\dag_{\alpha k\sigma}a_{\alpha k\sigma}+\sum_k(\Delta_\alpha e^{i\varphi_\alpha} a_{\alpha, k\downarrow}
a_{\alpha,-k\uparrow}+h.c.),\notag\\
H_D&=&\sum_{\sigma,j=1}^2\varepsilon_jd^\dag_{j\sigma}d_{j\sigma}+\sum_\sigma(t_cd^\dag_{2\sigma}d_{1\sigma}+h.c.)\notag\\
&&+\sum_{j=1}^2U_jn_{j\uparrow}n_{j\downarrow},\notag\\
H_T&=&\sum_{\alpha k\sigma}(V_{\alpha U}a^\dag_{\alpha k\sigma}d_{1\sigma}+V_{\alpha D}a^\dag_{\alpha k\sigma}d_{2\sigma}+h.c.).
\end{eqnarray}
\end{small}
$a^\dag_{\alpha k\sigma}$ ($a_{\alpha k\sigma}$) is an operator to
create (annihilate) an electron with momentum $k$ and spin
orientation $\sigma$ in SC-$\alpha$. $\varepsilon_{\alpha k}$
denote the corresponding energy levels. $\Delta$ is the
superconducting gap and $\varphi_\alpha$ is the phase of the
superconducting order parameter. $d^\dag_{j\sigma}$ ($d_{j\sigma}$) is an operator to create
(annihilate) an electron with energy $\varepsilon_{j}$ and spin
orientation $\sigma$ in QD-$j$, respectively. $U_j$ indicates the
strength of intradot Coulomb repulsion, and $t_c$ is the interdot
coupling coefficient. $V_{\alpha U(D)}$ denotes the coupling between SC-$\alpha$ and
QD-$1(2)$.
\par
In this work we would like to take a few simplifications to make
clearer the Josephson effect in such a SC/DQD/SC structure. The SCs are assumed to be identical
($\varepsilon_{\alpha k}=\varepsilon_k$ and $\Delta_\alpha=\Delta$) except for a finite phase difference
$\varphi_L=-\varphi_R=\varphi/2$ and $\varphi_M=0$.
The Josephson current between at
zero temperature can be evaluated by deriving the GS
energy $E_{GS}$ with respect to the superconducting phase
difference, i.e.,
\begin{equation}
I_J={2e\over \hbar}{\partial E_{GS}(\varphi)\over\partial\varphi}.
\end{equation}
\par
It should be emphasized that since we are interested in transport processes involving
Andreev reflection, we focus on the limit $\Delta\to \infty$. In this case,
as the quasiparticles in the SCs are inaccessible,
tracing out of the degrees of freedom of the SCs does not induce any dissipative dynamics in the DQD
system and it can be performed exactly. The resulting
Hamiltonian dynamics of the DQDs can be described by the
following effective Hamiltonian [20]
\begin{small}
\begin{eqnarray}
H_{\rm eff}&=&H_D\notag\\
&+&{1\over2}\sum_{\alpha}[\sqrt{\Gamma_{\alpha U}\Gamma_{\alpha D}}e^{i\varphi_{\alpha}}(d^\dag_{1\uparrow}d^\dag_{2\downarrow}+d^\dag_{2\uparrow}d^\dag_{1\downarrow})+h.c.]\notag\\
&-&{1\over2}\sum_{\alpha}[\Gamma_{\alpha U}e^{i\varphi_{\alpha}}d^\dag_{1\uparrow}d^\dag_{1\downarrow}+\Gamma_{\alpha D}e^{i\varphi_{\alpha}}d^\dag_{1\uparrow}d^\dag_{1\downarrow}+h.c.].\notag\\
\end{eqnarray}
\end{small}
$\Gamma_{\alpha U(D)}$, defined by $\Gamma_{\alpha U(D)}=\pi \sum_k |V_{\alpha U(D)}|^2 \delta(\omega-\varepsilon_k)$, describes the coupling strength between the DQDs and leads. Within such an approximation, the Josephson current between two SCs can be evaluated by diagonalizing $H_{\rm eff}$. It is known that in this structure, fermion parity (FP) is good quantum number. The Hamiltonian matrix should be written out according to FPs, namely,
\begin{small}
\begin{eqnarray}
&&H^{(e)}_{\rm eff}=\notag\\
&&\left[
\begin{array}{cccccccc}
  0& -A &  A & 0 &0 & A & A &  0  \\
  -A^*& 2\varepsilon &  0 & 0 &0 & -t_c & -t^*_c &  A  \\
   A^* & 0 & 2\varepsilon & 0 &0 &  t_c & t^*_c &  -A \\
    0 & 0 & 0& 2\varepsilon & 0 &0 & 0 & 0 \\
     0 & 0 &  0 & 0 & 2\varepsilon & 0& 0 &  0  \\
      A^* & -t^*_c &  t^*_c & 0 &0 & U+2\varepsilon& 0  &  A  \\
        A^* & -t_c &  t_c & 0 &0 & 0 & U+2\varepsilon &  A  \\
        0 & A^*& -A^* & 0 &0 & A^*& A^* & 2U+4\varepsilon
\end{array}
\right],\notag
\end{eqnarray}
\end{small}
and
\begin{small}
\begin{eqnarray}
&&H^{(o)}_{\rm eff}=\notag\\
&&\left[
\begin{array}{cccccccc}
  \varepsilon & 0 & t^*_c  & 0 &-A & 0 & A &  0  \\
  0 & \varepsilon &  0 & t^*_c &0 & -A & 0 &  A  \\
   t_c & 0 & \varepsilon & 0 &A &  0 & -A &  0 \\
    0 & t_c & 0& \varepsilon & 0 &A & 0 & -A \\
     -A^* & 0 &  A^* & 0 & U+3\varepsilon & 0& -t^*_c &  0  \\
      0 & -A^* &  0 & A^* &0 & U+3\varepsilon& 0  &  -t^*_c \\
        A^* & 0 &  -A^* & 0 &-t_c & 0 & U+3\varepsilon & 0 \\
        0 & A^*& 0 & -A^* &0 & -t_c & 0 & U+3\varepsilon
\end{array}
\right]\notag
\end{eqnarray}
\end{small}
with $\Gamma_{\alpha\beta}=\Gamma_0$ and $A=\Gamma_0(1+2\cos{\varphi\over2})$.
Following the theory in the above section, we proceed to discuss the property of the Josephson current between SC-$L$ and SC-$R$.
\par

In Fig.1(b) we define $E_g=E^{(e)}_{GS}-E^{(o)}_{GS}$ and present the difference between the different-FP GS energies. The structural parameters are taken to be $\varepsilon_j=\varepsilon$, $t_c=4$, and $U=10$, respectively. The result in this figure shows that in such a structure, the Kramers degeneracy of the GS energies in different GSs is clearly broken. Changing $\varepsilon$ and $\varphi$ can alter the GS parity, with  the transition position (i.e., position of $E_g=0$) denoted by the dotted lines. In addition, one can find that in most of the parameter space, the energy of the odd-parity GS is lower than the even-parity GS. Such a result is opposite to that in the normal SC-existed system where $E_g$ is always less than zero due to the presence of Cooper pairs [19]. The other phenomenon is that $E_g$ presents different oscillation manners in the regions $\varepsilon>0$ and $\varepsilon<0$. Following the result in Fig.1(b), we plot the spectra of the Josephson currents in different FPs, as shown in Fig.1(c)-(d). It is evident that both of the Josephson currents in different FPs oscillate in $4\pi$ period with the similar directions of them. However, the differences between $I_e$ and $I_o$ are easy to observe. In addition to their different magnitudes, the odd-FP current is suppressed completely in the region $\varepsilon<-7.0$. This result can be explained by solving equation $H^{(o)}_{\rm eff}|\Psi^{(o)}\rangle=E^{(o)}|\Psi^{(o)}\rangle$. Under the condition of the parameters in Fig.1, the eigenvalues of $H^{(o)}_{\rm eff}$ can be listed, i.e., $E^{(o)}_{1(2)}=6+3\varepsilon$, $E^{(o)}_{3(4)}=4+\varepsilon$, and $E^{(o)}_{5\uparrow(\downarrow)}=E^{(o)}_{6\uparrow(\downarrow)}=5+2\varepsilon\pm\sqrt{93+18\varepsilon+\varepsilon^2+16\cos{\varphi\over2}
+8\cos\varphi}$, respectively. Note that the double degeneracy of these eigenvalues originates from the spin symmetry of basis of $H^{(o)}_{\rm eff}$. Surely, with the decrease of $\varepsilon$, $E^{(o)}_{1(2)}$ have an opportunity to become the GS level. And then, its independence of $\varphi$ suppresses the Josephson current directly. As is known, the degeneracy of the different-FP GSs allows to define a pair of Majorana operators ($\gamma_1$, $\gamma_2$) which transform the two GSs into each other. Therefore, we ascertain that the Majorana modes can be realized in such a triple-terminal junction.

\par

Coulomb interaction is a key factor to influence the properties of Josephson current. As an extreme case, when $U$ increases to infinity, the double occupation in each QD will be forbidden, which will inevitably transform the Josephson effect. Next, we would like to discuss the Josephson effect by taking $U\to\infty$. In such a case, the dimension of $H_{\rm eff}^{(e)}$ will decrease to five, and its eigenvalues are $E^{(e)}_{1}=E^{(e)}_{2(3)}=2\varepsilon$, and $E^{(e)}_{4(5)}=\varepsilon\pm\sqrt{6+\varepsilon^2+8\cos{\varphi\over2}+4\varepsilon}$, respectively. On the other hand, $H_{\rm eff}^{(o)}$ will reduce to one $4\times4$ matrix with the eigenvalues $E^{(o)}_{1(2)}=\varepsilon+t_c$ and $E^{(o)}_{3(4)}=\varepsilon-t_c$. With these results, we can write out the expression of $E_g$ directly, i.e.,  $E_g=-\sqrt{6+\varepsilon^2+8\cos{\varphi\over2}+4\cos\varphi}+t_c$. Evidently, $E_g$ is a even function of $\varepsilon$. Its detailed property can be observed in Fig.2(a). Suppose $E_g=0$, the transition position of the GS parity can be well-defined with the relationship $\varepsilon=\pm\sqrt{t_c^2-6-8\cos{\varphi\over2}-4\cos\varphi}$. As for the Josephson current in the limit $U\to\infty$, it comes into being only in the even-parity case because of the $\varphi$ independence of $E^{(o)}$. As shown in Fig.2(b), the Josephson current oscillates in $4\pi$ period and its maximum shifts to the point of $\varepsilon=0$, though the current magnitude is somewhat weakened in comparison with the case of $U=10$.

\begin{figure}[htb]
\begin{center}\scalebox{0.032}{\includegraphics{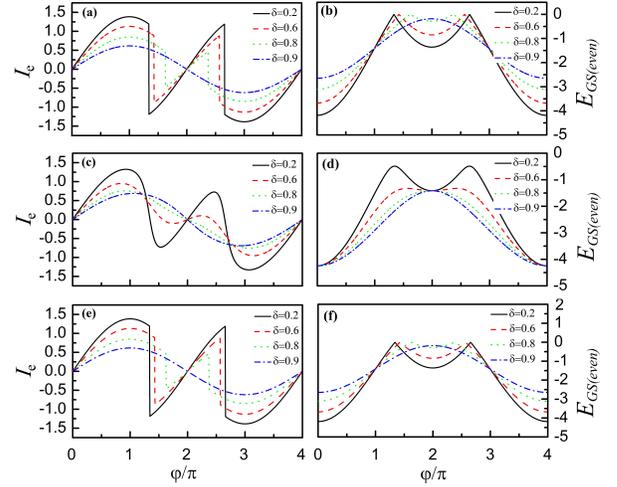}}
\caption{Even-FP Josephson current and GS level as functions of $\varphi$ in the asymmetric QD-superconductor coupling case. (a)-(b) Josephson current and GS level in the case of up-down asymmetry with $\Gamma_{LU}=\Gamma_{RU}=\Gamma_0+\delta$ and $\Gamma_{LD}=\Gamma_{RD}=\Gamma_0-\delta$. (c)-(d) Results of left-right asymmetry with $\Gamma_{LU}=\Gamma_{LD}=\Gamma_0+\delta$ and $\Gamma_{RU}=\Gamma_{RD}=\Gamma_0-\delta$. (e)-(f) Results of up-down left-right asymmetry with $\Gamma_{LU}=\Gamma_{RD}=\Gamma_0+\delta$ and $\Gamma_{LU}=\Gamma_{RD}=\Gamma_0-\delta$.
\label{Struct}}
\end{center}
\end{figure}
\par
The results in Fig.2 show that Majorana modes survive in the extreme case of $U\to\infty$. Due to the decrease of the dimension of $H^{(e)}_{\rm eff}$ in such a case, we are able to present the definition of the Majorana modes in an analytical way. The role of Majorana operator is to change the GS parity, which can be expressed as
\begin{equation}
\gamma_1|\Psi^{(o)}\rangle=|\Psi^{(e)}\rangle.\label{gamma}
\end{equation}
Taking the case of $\varphi=\pi$ as an example, where $\Psi^{(e)}_{GS}={1\over\sqrt{2+(\varepsilon+\sqrt{2+\varepsilon^2})^2}}[-(\varepsilon+\sqrt{2+\varepsilon^2})|00\rangle-|\downarrow\uparrow\rangle+|\uparrow\downarrow\rangle]$ and $\Psi^{(o)}_{GS}={1\over2}[|\uparrow0\rangle-|0\uparrow\rangle+|\downarrow0\rangle-|0\downarrow\rangle]$, the Majorana operator can be worked out as follows. First, we define the spinful Marjoana mode with its operator being
$\gamma_1=an_{1\downarrow}\gamma_{1\uparrow}+bn_{1\uparrow}\gamma_{2\downarrow}+cn_{2\downarrow}\gamma_{1\uparrow}+dn_{2\uparrow}\gamma_{1\uparrow}+
en_{2\uparrow}\gamma_{1\downarrow}+fn_{2\downarrow}\gamma_{1\downarrow}+g\gamma_{1\uparrow}+h\gamma_{2\downarrow}+jn_{1\downarrow}n_{2\uparrow}$ and $\gamma_{j\sigma}={1\over\sqrt{2}}(d^\dag_{j\sigma}+d_{j\sigma})$. Next, by using the relationship $\gamma_1=|\Psi^{(e)}\rangle\langle\Psi^{(o)}|$, the relevant parameters in the expression of $\gamma_1$ can be obtained. They are
$a=d=-g={2\sqrt{2}(\varepsilon+\sqrt{2+\varepsilon^2})\over\sqrt{2+(\varepsilon+\sqrt{2+\varepsilon^2})^2}}$,
$b=c={2\sqrt{2}(\varepsilon+\sqrt{2+\varepsilon^2}-1)\over\sqrt{2+(\varepsilon+\sqrt{2+\varepsilon^2})^2}}$, and $e={2\sqrt{2}\over\sqrt{2+(\varepsilon+\sqrt{2+\varepsilon^2})^2}}$ whereas $f=h=j=0$. It is certain that $\gamma_2$ has a similar
form with site indexes 1 and 2 interchanged. Based on the above analysis, we ascertain that in such a triple-terminal DQD junction, the Majorana mode can be constructed. Since the realization of Majorana mode is irrelevant to the magnetic factors and other fields, such a scheme is more feasible compared with the previous works.

\par
Since the parallel geometry of such a Josephson junction, the symmetry of QD-SC couplings can be adjusted to be various manners, such as up-down asymmetry, left-right asymmetry, and up-down left-right asymmetry. Different asymmetry manners will modify the Josephson effect in special ways. We next suppose $U\to\infty$ and discuss the influence of the asymmetry of the QD-SC coupling manner on the formation of Majorana mode. The results of the Josephson current and GS level are shown in Fig.3. It can be found that regardless of the QD-SC coupling manners, the increase of $\delta$ efficiently decreases the magnitude of the Josephson current, followed by its weakened oscillation near the position $\phi=2\pi$. At the asymmetric limit of $\delta=0.9$, the additional oscillation of $I_e$ around $\varphi=2\pi$ disappears, as a result, the Josephson current obeys the relationship that $I_e\sim\sin{\varphi\over2}$ [See Fig.3(a), Fig.3(c), and Fig.3(e)]. Such a result can be well understood based on the GS-level oscillation shown in Fig.3(b), Fig.3(d), and Fig.3(f). These results mean that the case of asymmetric QD-SC coupling can not destroy the Majorana mode but promote its appearance.

In conclusion, we have studied the Josephson effect in one triple-terminal junction with embedded parallel-coupled DQDs. As a consequence, it has been found that the inter-SC supercurrent oscillate in $4\pi$ period, following the adjustment of the phase differences among the SCs. Moreover, such a result is robust and independent of FPs, intradot Coulomb strength, and the QD-SC coupling manner. By introducing the concept of spinful many-particle Majorana modes, the analytical definition of the Majorana operator has been presented. In addition, we have found that the asymmetric QD-SC coupling manner is conducive to the realization of Majorana modes in such a system. In view of all the results, we consider that this system can be a promising candidate for the realization of Majorana modes in a nonmagnetic system.
\par
This work was financially supported by the Fundamental Research
Funds for the Central Universities (Grants No. N130405009 and No.
N130505001), the Natural Science Foundation of Liaoning province of
China (Grant No. 2013020030), and the Liaoning BaiQianWan Talents
Program (Grant No.2012921078).

\clearpage

\bigskip

\end{document}